# METIS: the Mid-infrared E-ELT Imager and Spectrograph

Bernhard R. Brandl*[a], Markus Feldt[b], Alistair Glasse[c], Manuel Guedel[e], Stephanie Heikamp[a], Matthew Kenworthy[a], Rainer Lenzen[b], Michael R. Meyer[f], Frank Molster[g], Sander Paalvast[h], Eric J. Pantin[i], Sascha P. Quanz[f], Eva Schmalzl[a], Remko Stuik[a,j], Lars Venema[a,k], Christoffel Waelkens[l], and the NOVA-Astron Instrumentation Group[j].

[a]Leiden Observatory, Leiden University, P.O. Box 9513, 2300 RA Leiden, The Netherlands;
[b]Max-Planck-Institut für Astronomie, Königstuhl 17, 69117 Heidelberg, Germany;
[c]UK Astronomy Technology Centre, Edinburgh EH9 3HJ, UK;
[e]University of Vienna, Department of Astrophysics, Türkenschanzstrasse 17, A-1180 Vienna, Austria;
[f]ETH Zürich; Institute for Astronomy, Wolfgang-Pauli-Strasse 27, CH-8093 Zürich, Switzerland;
[g]NOVA , Leiden University, P.O. Box 9513, 2300 RA Leiden, The Netherlands;
[h]Janssen Precision Engineering BV, Aziëlaan 12, 6199AG, Maastricht-Airport, the Netherlands;
[i]Groupe LFEPS, Service d'Astrophysique, CE Saclay DSM/DAPNIA/Sap, 91191 Gif sur Yvette Cedex, France;
[j]NOVA-ASTRON, P.O. Box 2, 7990 AA Dwingeloo, The Netherlands;
[k]ASTRON, P.O. Box 2, 7990 AA Dwingeloo, The Netherlands;
[l]Instituut voor Sterrenkunde, K.U.Leuven, Celestijnenlaan 200D, B-3001 Leuven, Belgium.

## ABSTRACT

METIS will be among the first generation of scientific instruments on the E-ELT.  Focusing on highest angular resolution and high spectral resolution, METIS will provide diffraction limited imaging and coronagraphy from 3-14µm over an 20″×20″ field of view, as well as integral field spectroscopy at R ~ 100,000 from 2.9-5.3µm.  In addition, METIS provides medium-resolution (R ~ 5000) long slit spectroscopy, and polarimetric measurements at N band.  While the baseline concept has already been discussed at previous conferences, this paper focuses on the significant developments over the past two years in several areas: The science case has been updated to account for recent progress in the main science areas circum-stellar disks and the formation of planets, exoplanet detection and characterization, Solar system formation, massive stars and clusters, and star formation in external galaxies.  We discuss the developments in the adaptive optics (AO) concept for METIS, the telescope interface, and the instrument modelling.  Last but not least we provide an overview of our technology development programs, which ranges from coronagraphic masks, immersed gratings,  and cryogenic beam chopper to novel approaches to mirror polishing, background calibration and cryo-cooling.  These developments have further enhanced the design and technology readiness of METIS to reliably serve as an early discovery machine on the E-ELT.

## 1. INTRODUCTION

METIS is the name of the 'Mid-infrared ELT Imager and Spectrograph', the only E-ELT instrument to cover the scientifically important thermal/mid-infrared wavelength range from 3 – 14 (20) µm.  In a nutshell, METIS will provide a new window to the thermal infrared Universe at angular resolutions similar to what HST achieves in the optical, and adds unique high resolution (R~100,000) integral-field spectroscopy with JWST-like sensitivity (for unresolved lines).

*brandl@strw.leidenuniv.nl; phone +31 71527 5830

METIS has already been described in several papers of this conference series (e.g., [1], [2]). Here we briefly repeat the basic concept, but the main focus of this paper is on specific areas with recent progress, spanning from science to technology. The main instrument specifications have been derived from the METIS Science Case but taking several, fundamental boundary conditions into account: (1) The E-ELT with its 39m aperture will provide unsurpassed angular resolution at high sensitivity. Hence, METIS shall focus on the tremendous gain in angular resolution with respect to other facilities. (2) ALMA is now in operation and the James Webb Space Telescope (JWST) will be launched in 2018, with a mission lifetime of 5 years or more. Both ALMA and JWST will provide the ideal sample for METIS follow-ups. Hence, METIS shall provide unique observing modes and capabilities, which are complementary to ALMA and JWST. (3) Cerro Armazones, where the E-ELT will be located, is an excellent in terms of seeing and clear nights, but not necessarily in terms of precipitable water vapor. Hence, METIS shall be optimized for those wavelength bands, which provide the highest sensitivity for the given site, and pay careful attention to background and spectral calibrations. (4) Given the large thermal background from atmosphere and telescope on the ground, METIS will have a relatively low surface brightness sensitivity and shall focus on relatively compact objects or structures, which are close to the diffraction limit. (5) In order to provide a reliable workhorse instrument for the E-ELT in its first phase, METIS shall be based on a concept of low complexity and risks, requiring high technology readiness levels (TRL).

Programmatically, METIS has now been clearly defined in the E-ELT instrument roadmap as an early instrument (#3 in the sequence of planned commissionings). In early 2014, the E-ELT Project Science Team (PST) has defined the Top Level Requirements (TLRs), which have subsequently been approved by ESO's Scientific Technical Committee (STC). ESO had already previously concluded to negotiate the contract for a mid-infrared instrument directly with the METIS consortium, and the METIS consortium will soon start a so-called Interim Study, which further advances the status of the technical components and telescope interface, as well as the preparations of the Statement of Work (SoW), and other managerial documents, which will be required before phase B. At the time of writing, the kick-off of the METIS phase B will be in late 2015 with an anticipated commissioning at the E-ELT around 2025.

The METIS consortium consists of seven institutions: NOVA/Leiden University (the Netherlands, PI: B. Brandl), MPIA (Germany, co-I: M. Feldt), CEA-Saclay (France, co-I: E. Pantin), UK-ATC (United Kingdom, co-I: A. Glasse), KU Leuven (Belgium, co-I: Ch. Waelkens), ETH Zürich (Switzerland, co-I: M. Meyer), and the Universität Wien (Austria, co-I: M. Guedel). The METIS consortium draws heavily from its long-term successful experience with numerous ground- and space-based infrared instruments, most recently MIRI for the JWST.

The work presented in this paper is based on important contributions from all members of the METIS team, many more than could be listed as co-authors. This paper is organized as follows: Highlights of the METIS science case will be summarized in section 2, followed by the instrument overview in section 3, which includes not only an instrument baseline description but also the interface to the telescope and adaptive optics. In section 4 we discuss the technology advances, recent developments and prototypes of key components of METIS, which we have been performing over the past years to enhance the TRL of METIS. Finally, section 5 provides a summary.

## 2. SCIENCE CASE

It is obvious that the E-ELT will open up new parameter spaces for optical/infrared astronomy and enable observations in the thermal/mid-IR range, which have never been possible before. METIS' unique contributions to astrophysics in the 2020s will likely be in science areas where high spatial or high spectral resolution or a combination of both is crucial. It is important to keep in mind that METIS will be a general purpose science instrument. The METIS science case – as conceived more than 10 years before first light – covers a broad range of science topics, and a comprehensive description, even a complete listing, is beyond the scope of this paper. Instead, we focus on two areas, where METIS will play a key role: circum-stellar disks and extrasolar planets. Three additional topics (Solar System formation, massive stars, and star formation in galaxies) are briefly discussed in the second part to illustrate the breadth of science areas METIS will contribute to.

## 2.1 Circum-stellar Disks and the Formation of Planets

A central challenge in understanding the nature of exoplanets is to link the observed diversity of mature planetary systems to the observed properties of the gas-rich proto-planetary disks from which they form. What are the necessary conditions for planet formation, and how does the initial star and disk structure / composition determine the properties the final planetary system? A critical environment is the 1-10 AU region, where nearly all planets (both gas giant and rocky, terrestrial) are thought to form. Pioneering studies of disk structures have been made using long baseline interferometry with the VLTI and will continue with future instruments (e.g., MATISSE). However, METIS, on the filled aperture E-ELT, will be orders of magnitude more sensitive and will provide information on the disk dynamics thanks to its high spectral resolution. METIS will offer imaging at unprecedented spatial resolution of lines from gas-phase species and thermal emission from hot dust at great sensitivity, enabling study of low surface brightness disk structures as small as 1 AU across, complementary to ALMA and JWST. METIS observations are likely to be transformative in our understanding of all stages of planet formation. Key topics include: (1) Compositional analysis of gas in planet-forming disk regions; (2) detection and direct characterization of young, forming proto-planets; (3) spectral imaging of ices and PAHs in disks; (4) resolved imaging of debris disks; and (5) search for direct signatures of ongoing terrestrial planet formation (e.g., proto-planet collision afterglows).

The unique combination of high spectral and spatial resolution offered by METIS allows spectral imaging of the spatial distribution of disk material and kinematics of warm proto-planetary gas. Using CO rovibrational emission from the fundamental band at 4.7 μm as a ubiquitous, easily observable and well-studied tracer, METIS will resolve moving gas structures as small as 2 AU in size at typical distances of 100 pc and as little as 1 AU for the closest disks. This emission traces gas at a few hundred K to about 1000 K. The transitions are excited by UV fluorescent excitation [3] and by IR resonant pumping [4]. This leads to strong line emission from small disk radii, and lower, but significant, extended emission from several to 10s of AU, depending on the strength and hardness of the stellar spectrum. The gas traces the surface of the disk, which is optically thick in the vertical direction, but mid-IR data may penetrate close to the mid-plane once the gas dissipates or is depleted by the dynamic action of a planet [5]. Already with current facilities, the infrared line emission is detected in the majority of disks and has been spatially resolved in a few of them, in particular those with partially cleared-out inner regions, e.g. [6]. Typically the CO emission extends over only a couple of spaxels in favorable cases on 8-10 m telescopes. The 4-5 times higher spatial resolution with the ELT will, in combination with higher surface brightness sensitivity, lead to spectrally resolved CO line cubes over as many as 50 spaxels or more.

In addition, the high sensitivity of METIS may even allow the detection of spatially unresolved, warm gas and dust in the immediate vicinity of newly formed giant proto-planets of a few Jupiter masses [7]. Thick, circum-planetary disks surround giant planets during the proto-planetary disk phase. In particular, METIS will be able to detect warm CO gas in local Keplerian motion around the planet. A key property of the gas signature from a circum-planetary disk is that it will be offset from the local disk velocity by up to 5-10 km/s due to the planet's own motion around the star, thus avoiding interference from ambient disk gas. Figure 1 shows the results from simulated METIS observations illustrating both the detection of circum-stellar CO emission and of circum-planetary CO emission of a newly formed gas giant proto-planet.

## 2.2 Exoplanet Detection and Characterization

Today more than 1,500 confirmed exoplanet are known [9]. The majority are planets detected by radial velocity (RV) or transit observations, yielding information on exoplanet masses, orbital periods, and bulk densities (and hence average compositions). By mid 2020, tens of thousands of exoplanets will be known. GAIA alone is expected to detect between 10,000 and 50,000 exoplanets by tracing the astrometric reflex motion of the host star, e.g. [10]. METIS will be a key instrument to provide direct imaging follow-up observations of the most suitable targets detected by RV, transit or astrometric observations to determine their luminosities and characterize their atmospheric properties and dynamics. In general, high-contrast imaging at 3-5 μm on the E-ELT will be a powerful capability for finding gas giant exoplanets beyond 3 AU. METIS is capable of detecting 4-5 times more planets than NACO on the VLT for a comparable L-band survey of 100 nearby young FGK stars. Moreover these planets will be mostly closer to their host stars (< 20 AU) and factors of several lower in mass, reaching below one Jupiter mass for some targets. In this context it is worth mentioning that most of the directly imaged exoplanets at large separations known today were found in the L-band, e.g., Beta Pic b [11] and HD95086 b [12].

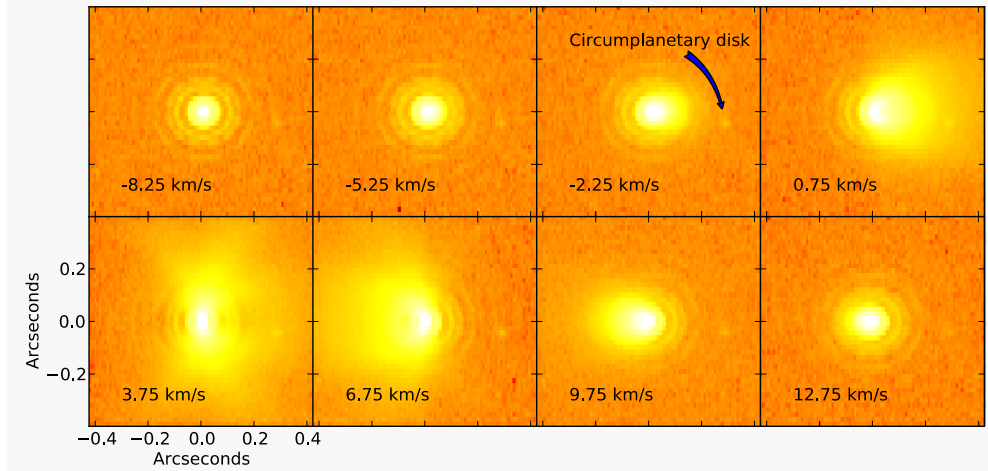

Figure 1 Simulation of a 1 hour METIS-IFU observation of a gas-rich protoplanetary disk around a young, solar-mass star with an embedded protoplanet. Offset by 30 AU from the central star, and indicated by the arrow, is a 10 $M_{Jup}$ protoplanet surrounded by a circumplanetary disk. The presence of a compact object is demonstrated by broad line emission at the location of the protoplanet; the warm circumplanetary disk is clearly separated from the cooler disk material in several planes of the cube. The high signal-to-noise of the ambient CO disk emission, seen at the less extreme velocities, further shows that the gas structure of inner disks can be imaged with very high fidelity at scales as small as 5 AU. The spectral setting is centered on the 4.7 micron CO rovibrational band and shown is a stack of 5 CO lines (observable in a single METIS setting). The CO lines are matched to existing observations with e.g., VLT-CRIRES of similar disks. Continuum emission has been subtracted from the cubes, so that all emission is from CO gas. The underlying model is based on non-LTE two-dimensional disk models calculated by RADMC and RADLite [8].

In addition to gas giant planets, METIS might also be able to detect smaller planets in the solar neighborhood. Dedicated surveys for terrestrial exoplanets are already ongoing or planned for the near future, and aim at detecting rocky planets by methods of transits (e.g., MEarth, TESS, PLATO), RV (e.g., CARMENES, ESPRESSO) or astrometry (e.g., GRAVITY). While not all efforts will likely be successful, there is a high probability that at least one of these methods is going to reveal one or more nearby terrestrial exoplanets. Determining the albedo or the chemical composition of the atmosphere of rocky exoplanets will remain challenging, but would be of utmost scientific interest. As an example, we consider an Earth twin orbiting Alpha Centauri A in a Venus-like orbit with a semi-major axis of 0.72 AU. Such a planet would have an angular separation of up to 540 marcsec from its host star, and a flux of 0.04 mJy in the N-band. METIS would be able to directly detect such a planet within only 1hour exposure time at 6-σ (Figure 2).

Furthermore, the high-dispersion integral field spectrograph provides a unique capability to study exoplanet atmospheres. Already today, dayside and transmission spectroscopy using CRIRES on the Very Large Telescope traces molecules in exoplanet atmospheres, such as water and carbon monoxide and also directly tracks orbital velocities and inclinations of giant exoplanets, e.g. [13], [14]. Today's measurements, though, are limited to studying integrated signals originating in molecular bands. METIS' strong increase in sensitivity compared to present-day facilities promises to open a whole new range of measurements. For the first time we will be able to measure the strengths of individual lines, thus directly probing the temperature-pressure profile. METIS observations will also reveal detailed signatures of planet rotation and circulation, which show up as line broadening and velocity-shifts during ingress and egress of a transit. Another aspect of high-resolution dayside spectroscopy is that a planet signal can be monitored along a large part of its orbit, thus tracing seasonal as well as rotational changes in molecular signals (from the morning-side to the evening-side of the planet, revealing possible photochemical processes or variations in the atmospheric temperature structure).

All these examples illustrate the enormous discovery potential for METIS, providing unique information on exoplanets and their atmospheres inaccessible to any other currently planned facility.

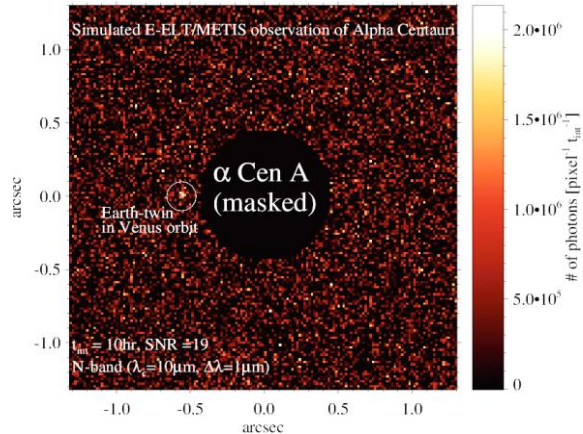

Figure 2 Simulated 10 h N-band coronagraphic observation of Alpha Centauri A. An Earth-twin (same radius and atmospheric composition, but with an average temperature of about 330K) in a Venus-like orbit would be detectable. The simulation assumes a sky brightness in N-band of -4.5 mag/arcsec$^2$, and an overall system throughput of 10%.

**2.3 Additional High-Impact Science**

Here we briefly touch upon three additional science areas, acknowledging that METIS' contribution to other exciting science topics like evolved stars, brown dwarfs, the Galactic Center, the growth of supermassive black holes in galaxies out to 100 Mpc, and the interstellar medium is being discussed elsewhere. Given the enormous discovery space of METIS compared to previous ground-based mid-IR instruments, the greatest discoveries METIS will likely make are those we cannot even imagine yet.

**Solar system formation**

The ages of the oldest analyzed solids in our Solar System are 4.57 billion years [15]. METIS, with its powerful combination of unprecedented spatial resolution and spectral mapping capabilities (at a range of spectral resolutions) will make fundamental contributions to discerning patterns in compositional gradients in Solar System bodies, providing key constraints on models of their formation and evolution. For example, trace species like $PH_3$ and CO in the troposphere of Uranus and Neptune, measured with high-dispersion in the 4-5 μm wavelength region, provide unique constraints on the internal mixing of these planets and thus their internal energy budgets which are dominated by formation processes.

**Massive stars and stellar cluster formation**

In many ways, massive stars ($\geq 8\ M_\odot$ and $\geq 10^3\ L_\odot$) shape the visible Universe. METIS on the E-ELT will enable vital progress in answering key questions related to massive star formation. Its wavelength range and spatial resolution will provide the capability to penetrate high extinction columns, measure the spatial scales, and detect the spectroscopic tracers associated with massive star formation, resolving individual massive young stellar objects (MYSOs) within star-forming regions throughout the Milky Way and clusters in nearby galaxies. While ALMA – being most sensitive to cool molecular gas – will provide constraints on the physical conditions from which the stars form, METIS – most sensitive to warm ionized or molecular gas – will enable characterization of the properties of the resulting proto-stars and clusters.

A key science project for METIS will be to image the youngest embedded clusters within Galactic massive star-forming regions with sufficient spatial resolution to separate the individual young stars and measure the radial dependence of the luminosity function. Using a sample of such regions, it would also be possible to search for variations in the star-forming rate and efficiency, vital inputs into semi-analytic models of galaxy formation and evolution.

**Star Formation in External Galaxies**

Understanding how galaxies formed and evolve remains one of the most challenging issues in modern cosmology. Over the past few years, tremendous progress was made in constraining at kilo-parsec scales the nature of the physical processes that drive the build-up of stellar mass across cosmic times. However, on the intermediate sub-galactic scales, the local conditions and processes that compressed the gas and triggered the formation of new stars at such high rates are still very uncertain. Combining the long wavelengths ($\geq 3$μm) and high angular resolution provided by METIS, we will be able to unveil the nature of the physical mechanisms at work in such environments, studying in detail the formation of dense and massive star clusters found in nearby luminous infrared galaxies and circum-nuclear starbursts. For illustration, Figure 3 shows the circum-nuclear starburst ring in NGC 7552, a face-on galaxy at approximately 20 Mpc distance, as observed with different instruments and telescopes (Spitzer, VLT) at different wavelengths and angular resolutions.

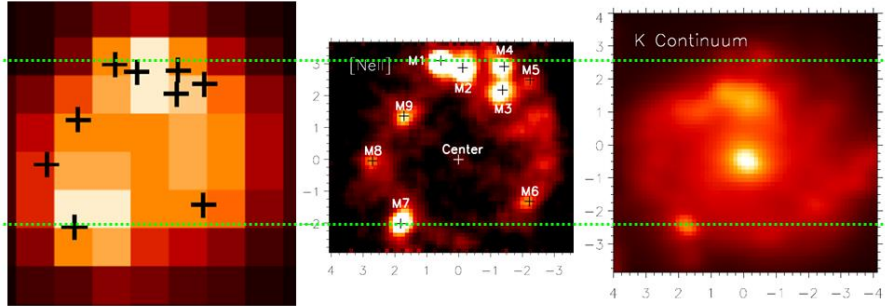

Figure 3: Infrared imaging of the circum-nuclear starburst ring in NGC 7552. Left: Spitzer-IRAC at 8 μm; center: VLT-VISIR [Ne II] at 12.8μm; right: VLT-SINFONI K-band (~2 μm). The black crosses indicate the location of the most mid-IR luminous clusters; the dotted horizontal lines have been added to guide the eye. Note that the K-band continuum emission does *not* coincide with the location of the most luminous clusters [16].

## 3. INSTRUMENT OVERVIEW

In this section we describe METIIS at the instrument level: the instrument baseline and conceptual design, its associated adaptive optics system(s), the interface to the telescope, and some results from performance modeling.

### 3.1 Instrument Baseline

The METIS baseline specification, as derived during the phase A study, includes two main subsystems as follows:

A **diffraction limited imager** at L/M and N band, with an approximately 20″×20″ wide FOV. The imager also includes the following observing modes:
- coronagraphy at L/M and N-band
- medium-resolution (900 ≤ R ≤ 5000) long slit spectroscopy at L/M and N band
- polarimetry at N-band.

An **IFU-fed, high resolution spectrograph** at L/M band [2.9 – 5.3μm] with an IFU field of view of about 0.4″×1.5″, and a spectral resolution R ~ 100,000.

With this baseline, METIS incorporates important capabilities of several successful VLT instruments, most notably NACO, CRIRES and VISIR. Figure 4 shows the phase space of spectral resolution and wavelengths, occupied by METIS in comparison with these VLT instrument capabilities.

Generally, ESO reevaluate the specifications for a given E-ELT instrument before phase B. The first step in this process is the definition of the TLRs by the PST. In 2014, the PST added two instrument modes to the TLRs as derived from the METIS phase A study: imaging and medium resolution slit spectroscopy at Q-band, and high resolution integral field spectroscopy at N-band. The addition of Q-band imaging and/or N-band IFS would add risk, cost and complexity with respect to the current baseline, and adds noticeably to the instrument budget. A more detailed study of the implications, the technical solutions and the trade-off optimization is the subject of the ongoing "Interim Study". However, some remarks can already be made at this point:

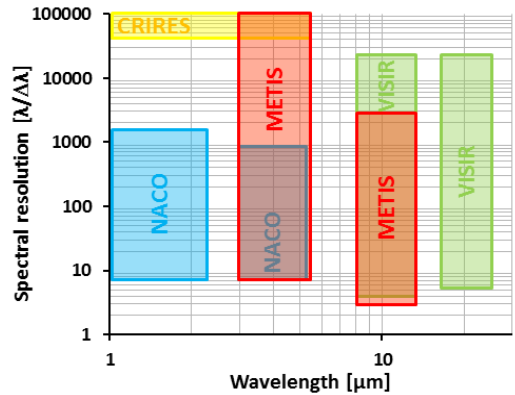

Figure 4 The wavelength/spectral resolution coverage of METIS in comparison with existing VLT instruments.

Q-band ($\lambda > 17$ μm) is scientifically very interesting for the study of brown dwarfs, Solar system objects, the inner structure of cool proto-planetary disks and extragalactic transients ([FeII] 17.99 μm line). Most conventional mid-infrared instruments also cover, at least partially, the Q Band as their Si:As BIB detectors are also sensitive in this wavelength range. However, the challenge for METIS is the dependency on AO correction. Even for a NIR WFS, the bandwidth required for both AO sensing and science Q-band channel is approximately a factor 20 in wavelength. Although most optical elements in METIS are reflective (achromatic), the cryostat window and the AO beam-splitter are not. Suitable materials which provide excellent transmission from $1 - 20$ μm *and* sufficient mechanical stability, optical quality, and size to serve as cryostat windows do not exist, and neither does a dichroic beam-splitter with excellent reflection between $1 - 2.5$μm and excellent transmission from $2.5 - 20$ μm. A possible solution may be the addition of a window exchange mechanism and a retractable AO beam-splitter.

Similarly, an N-band IFS would be scientifically highly desirable: The N band contains many unique spectral diagnostics, which are important for the proto-planetary disk chemistry at 1-10 AU; the [Ne II] line is typically 4.5 times narrower than the hydrogen lines and allows detailed studies of gas velocities around massive YSOs and in local ($z < 0.1$) galaxies. However, an N-band IFS would significantly increases the instrument size, require a diffraction-limited, large Echelle grating, and drive the cryo-cooling budget.

## 3.2 Conceptual Description

The conceptual design of METIS has been previously described ([1], [2]) and is built around a reimaging fore-optics module which includes a cold calibration unit, a cold pupil stop, a cold chopping mirror, and an image derotator (Figure 5). The imager and spectrograph modules are attached to the fore-optics and largely independent subsystems. The focal plane of the telescope lies outside the cryostat, as well as the warm calibration unit for easy access. The light enters through the entrance window in the top left of Figure 5. The first element in the common fore optics is the dichroic AO pick-off mirror, which directs the near-IR light to the AO wavefront sensor. A cold pupil stop is located at the intermediate pupil, together with a 2D-steerable mirror (section 4.6), and followed by the image derotator, after which the beam gets split into the different subsystems. The optical system uses all-reflective optics to accommodate testing and integration and to minimize chromatic aberrations.

The input to the L/M band high resolution IFU spectrometer pre-optics is picked off with a small mirror mounted on the wheel that also hosts the field masks for the imager, and transferred to the IFU. The pre-dispersion is accomplished by a ZnSe prism, the main dispersion by a large immersed Echelle grating. Tilts of the grating allow selecting the spectral range within the preselected diffraction order.

Retracting the pick-off allows the full field to enter the imager module. The L/M band and N band imager modules are very similar and contain a reimaging system with filter and grism wheels, pupil imaging optics, and a single detector each. For low-medium resolution slit spectroscopy a grism is inserted into the collimated beam and a slit mask is placed in the entrance focal plane.

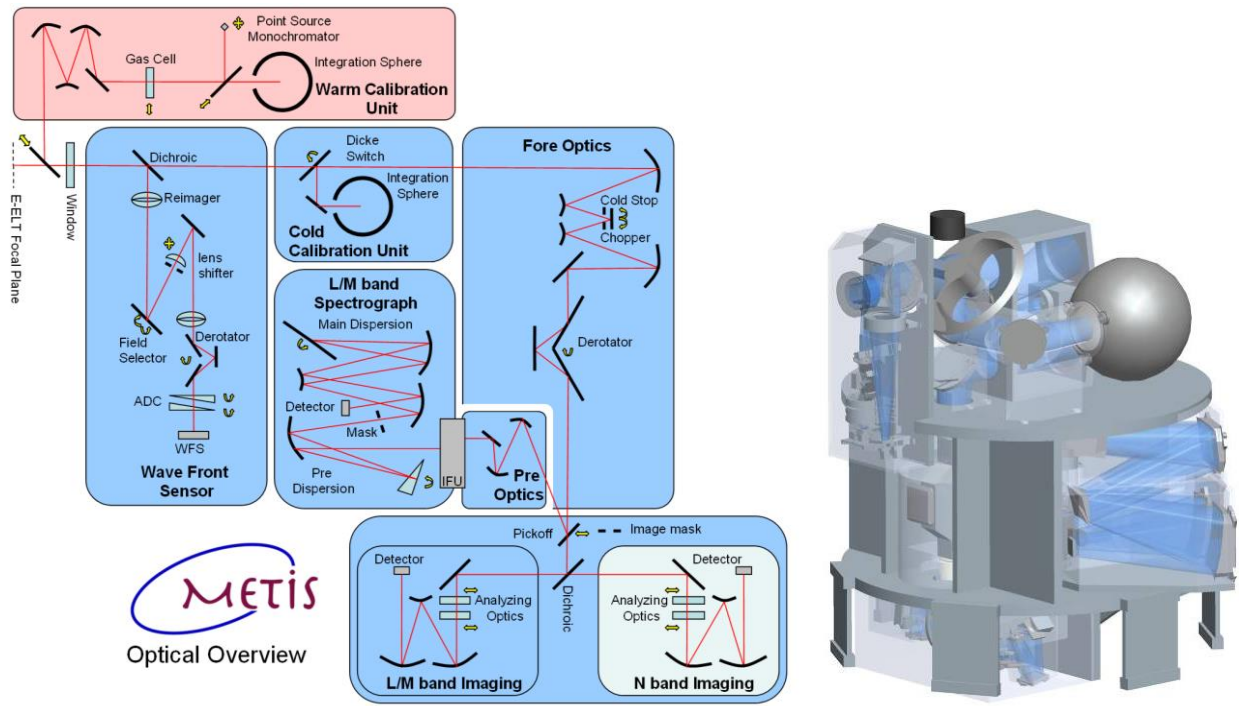

Figure 5 Left: Optical concept of METIS. Right: Outline of the optical system inside the METIS cryostat within the backbone structure. The opto-mechanical key components are also shown, including the Dicke switch (large tilted wheel on top), and the integrating sphere (top right) of the cold calibration unit. The two imager channels are at the bottom, the L/M band IFS to the right.

With only four focal plane arrays, the detector needs for METIS are modest: one 2048 × 2048 pixels HAWAII-2RG (Teledyne) array for L/M band imaging, one 4096 × 4096 pixels HAWAII-4RG (Teledyne) array for L/M band IFS, one 1024 × 1024 pixels AQUARIUS (Raytheon) array for N band imaging, and one 600 × 600 pixels SAFIRA (SELEX) APD for NIR wavefront sensing.

Figure 5 (right) shows the optical system inside the METIS cryostat within the backbone structure. The complete cold bench is divided in sub-modules, which will be designed, manufactured and tested as separate entities. The common backbone structure acts as mechanical and thermal interface and provides the alignment of the modules with respect to each other and the telescope image plane. Within the cold system different temperature levels have to be maintained. The detectors require approximately 40 K and 8K for the L/M band and N band, respectively. The N-band imager has to be cooled to below 30 K. The total mass of the cold system is 550 kg. The cold system is surrounded by a spherical cryostat, with a diameter of approximately 2.5m. For maintenance and service the lids provide easy access.

### 3.3 Adaptive Optics

All METIS observing modes require adaptive optics correction. Only under very favorable (and unpredictable) atmospheric conditions, METIS might be able to achieve quasi-diffraction-limited images at 10 μm even without any AO correction, provided the image quality is not too much degraded by telescope vibrations.

The conceptual design of the METIS AO system [17] is based on three boundary conditions: (i) the relatively low required complexity of the AO system (full AO correction at 5 μm requires only about $1/45^{th}$ of the number of actuators needed at 1μm), (ii) the relatively small required field of view, which is limited by the isoplanatic angle at λ = 3 μm in good agreement with the relatively low space density of interesting mid-IR targets, and (iii) that the METIS targets which require the highest AO performance (exoplanets and proto-planetary disks) are rather compact, largely self-referencing, and quasi on-axis.

**Internal SCAO**

The design of the internal Single Conjugate Adaptive Optics (SCAO) system is driven by the exo-planet science case, where-by definition-the target is bright and can be used as a reference object for the AO. The requirements on performance are therefore mainly on obtaining a high contrast in the region between 2 and 5 λ/D and low residual tip-tilt jitter to allow coronagraphy. The minimum Strehl Ratio that the AO system is expected to deliver, under median seeing conditions (seeing at zenith = 0.65″), at least 93% Strehl (goal: 95%) at 10μm, and at least 60% Strehl at 3.5μm, for zenith angles $\zeta \leq 45$ deg and an unresolved natural guide object with $m_K \leq 12$. This performance should be readily obtainable according to our simulations (Figure 6). Furthermore, the residual stability of the PSF centroid is expected to better than 0.003″ RMS over at least 15 minutes at zenith angles larger than 5 degrees.

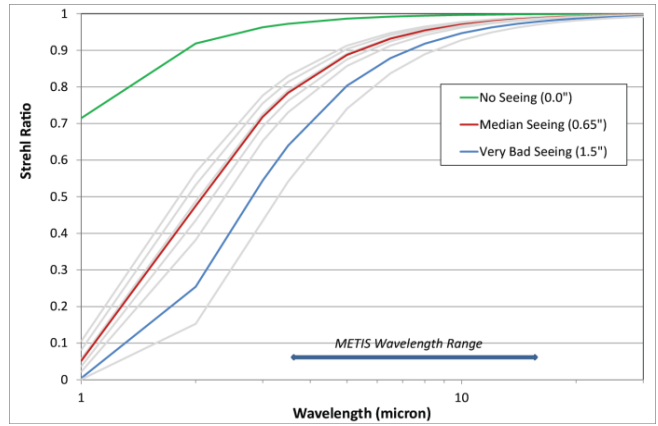

Figure 6 AO performance of the METIS internal SCAO system. Shown are the Strehl versus wavelength under various seeing conditions (0, 0.2, 0.4, 0.6, 0.65, 0.8, 1, 1.5, 2, and 2.5 arcseconds). Highlighted are only the 'no seeing' (basically only internal wavefront errors), 0.65″ and 1.5″.

The AO system that achieves these requirements is a classical AO system, albeit based on state-of the art components and using the M4 and M5 mirrors of the E-ELT for correction. As baseline, METIS will use a pyramid wavefront sensor (WFS), using 1-meter sub-apertures, gain control and a SELEX SAFIRA detector. This combination of wavefront sensor type, sub-aperture size and detector is already available, and the detector will soon demonstrate its performance in the VLTI instrument GRAVITY. The system is further complemented by an ADC, a field selector allowing the pick-up of the guide star anywhere in the METIS field of view (FoV). The WFS has been placed inside the cryostat to avoid having a warm beam splitter or dichroic in front of the instruments and close to the focal plane, which would substantially add to the thermal background and calibration challenges. As not all components may be readily cryo-compatible and a cooled pyramid WFS may be particularly challenging, this may add a substantial risk, which we are carefully investigating. The fallback solution is an IR Shack-Hartmann WFS.

An additional functionality requirement of the METIS internal SCAO system is that it shall be able to serve as a low order ("tip-tilt") WFS for the LTAO system (see below).

**External LTAO**

In order to achieve the full scientific potential of METIS on extragalactic sources or faint ($m_K > 12$), Galactic targets, laser guide stars (LGS) are needed for good sky coverage. LTAO would also provide a more uniform, and on average better, performance across the field of view. Initial simulations have shown that the LTAO system performance may be on-par with the SCAO system. However, since most observations, which require high contrast, can be done with the SCAO system, the requirements on LTAO performance have been reduced to a Strehl ratio of $\geq 75\%$ and $\geq 13\%$ at N- and L-band, respectively, over more than 80% of the sky.

In order to reduce residual motions between the external LTAO WFS and METIS, the METIS internal SCAO system can be reconfigured (decreased to a 2×2 or 3×3 subaperture system) to operate as low-order wavefront sensor for the LTAO. In this case the additional components required for LTAO are only the LGS WFS and the real-time control computer (RTC). In order to minimize the thermal background which METIS sees, the LGS need to be pick-up outside the METIS field of view, preferably with an annular mirror that keeps the science field of METIS clear of any additional optics. The

LTAO performance at METIS-wavelengths does not strongly depend on the off-axis distance of the LGS, which makes such a solution practical. An open question still is whether additional natural guide star sensors are needed around the METIS science field in case no sufficiently bright natural guide star is available in the patrol field of the internal sensor.

### 3.4 Telescope Interface

Since the end of the instrument phase A study, the interface between telescope and instruments has been revisited. This process was driven by the overall changes in the telescope design (e.g., smaller primary mirror, smaller instrument platforms) and the specific requirements by the individual instruments. Also, the design, implementation and responsibility for the laser guide star wavefront sensing required careful reconsideration. It is clear that in this phase of the project, many interface specifications will continue to change. The changes which occurred so far – telescope f-ratio, field curvature, telescope exit pupil location – do not affect the METIS conceptual design. The optical design can easily be adapted to accommodate these differences.

The integration of the METIS with the LTAO system has impact on the instrument architecture. The interface of instrument towards telescope needs to consider the space requirements of the telescope wavefront sensors, space requirements of the instrument configurations on the Nasmyth platform, costs of common infrastructure versus instrument specific infrastructure, programmatic and technological risks, etc. In the current concept, METIS receives light from the telescope without any additional mirror after M5 of the telescope. In front of METIS, a large annular mirror will reflect the light of the lasers outside the METIS science field to the LGS wavefront sensors. For the optimal performance, the laser guide stars will be projected on a circle with a radius of 1 – 1.25 arcminutes around the science target.

### 3.5 Performance Modelling

**METIS Instrument Simulator**

The METIS simulator [18] is a software package, written in the interactive language IDL. It is composed of five basic functional blocks, which together propagate the simulated image of an object above the atmosphere all the way to the detector readout electronics. These functional blocks are:
- The image track, which follows the underlying spatial image through the atmosphere, telescope, and METIS optics up to the camera or spectrometer.
- The throughput track, which calculates the effect that those optical components have in reducing the brightness of that image (independent of field position but as a function of wavelength).
- The background track, which is similar to the throughput track, but includes the thermal emission of all components.
- The camera or spectrometer function, which maps the received signal onto the relevant detector array.
- The detector function, which derives the photocurrent, detector noise, nonlinearity, saturation, and other detector artefacts.

The METIS simulator has become an invaluable tool to simulate realistic science observations and evaluate the discovery potential of METIS, in particular for marginally extended objects or complex source morphologies. For illustration of a typical simulated data product from the IFS, see Figure 7.

**Instrument Sensitivities**

The imaging mode of the simulator has also been used to derive the end-to-end instrument sensitivities. We have taken the specifications of the E-ELT telescope parameters as from the 2012 design report, assuming a total telescope emissivity of 15%, and the site information on Cerro Armazones from the TMT site study [19]. The atmospheric models were generated with the ESO SKYCALC Sky Model Calculator online software [20].

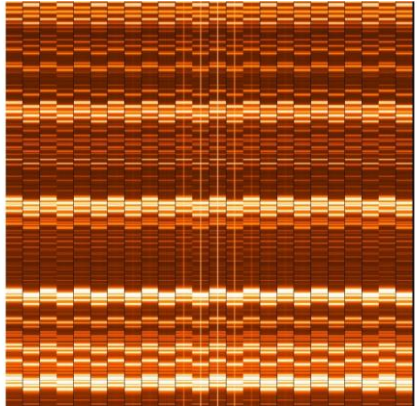

Figure 7: Detector output from a simulated one hour observation of a 10 mag star at L band as seen with the METIS high resolution IFS. The source can be seen as vertical lines on the central slices. Horizontal lines and bands are due to atmospheric emission.

The results of these estimates are shown in Figure 8. Although the *imaging* sensitivity of ground based IR instruments is inferior to space the huge gain in angular resolution makes more than up for this. Due to the large dispersion, the *spectroscopic* sensitivity of METIS is even comparable with JWST-NIRSPEC for unresolved spectral lines (usually the case for Galactic targets).

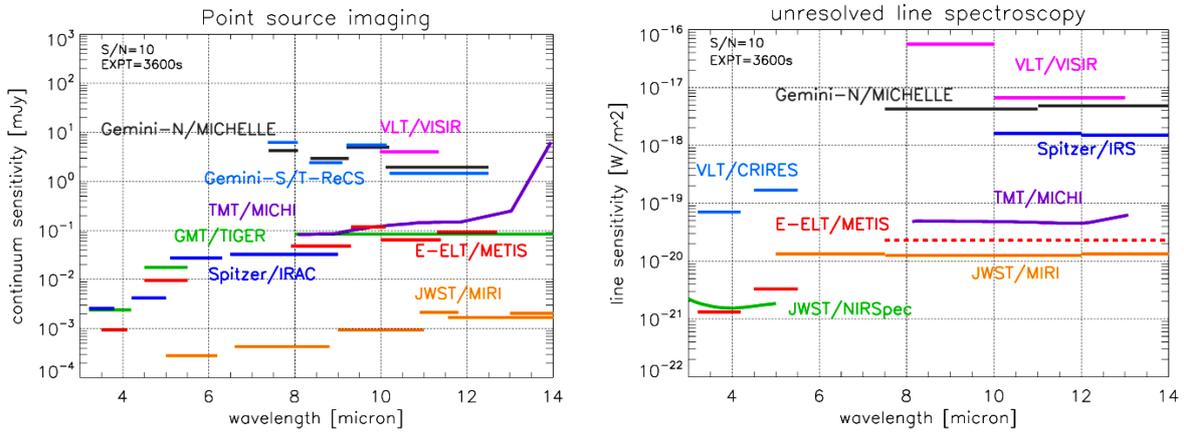

Figure 8 METIS point source sensitivities for a 10σ detection within one hour exposure time including 20% overhead for broadband imaging (left) and spectroscopy (right). Also shown are the published sensitivities for other facilities, corrected for the same integration time and significance of detection.

A more difficult case to accurately calculate (and specify) is the sensitivity to extended emission. Generally speaking, for extended emission the signal-to-noise (S/N) per pixel of an instrument which critically-samples the diffraction limit is independent of the telescope diameter, but most extended science targets possess internal structure with compact emission. The simulator is an invaluable tool to provide a realistic estimate of what METIS will be able to "see" on such complex targets.

**Saturation Limits**

We calculated saturation limits for both point and extended sources. The extended sources were simulated as uniform sources with a flat spectrum. The values given in Table 1 are for a full frame read-out time as a minimum integration time. So-called windowing is possible for both HAWAII-2RG and AQUARIUS science detectors, which allows for shorter integration times and higher saturation levels than given in Table 1.

Table 1 METIS saturation limits for point and extended sources for full frame read-outs, requiring a minimum integration time of 13 ms at L/M-band and 7 ms at N-band. For the modelled AO performance we assumed 0.6″ seeing, 30° zenith angle, and a guide star magnitude of $m_K$=10.

| Source type: | Point | | Extended |
|---|---|---|---|
| **Filter band [µm]** | [Jy] | [mag] | [MJy/sr] |
| $L_{wide}$, $\lambda_c$=3.6, FWHM=0.98 | 0.078 | 8.88 | 5.28 E6 |
| $L_{NB}$, $\lambda_c$=3.6, FWHM=0.02 | 3.75 | 4.67 | 1.46 E8 |
| $M_{wide}$, $\lambda_c$=4.8, FWHM=0.60 | 0.26 | 6.99 | 9.36 E6 |
| $M_{NB}$, $\lambda_c$=4.7, FWHM=0.02 | 11.0 | 2.93 | 4.43 E8 |
| $N_1$, $\lambda_c$=8.6, FWHM=1.4 | 43.9 | 0.11 | 4.30 E8 |
| $N_2$, $\lambda_c$=10.7, FWHM=1.4 | 47.0 | -0.44 | 4.71 E8 |
| $N_3$, $\lambda_c$=12.0, FWHM=1.4 | 53.0 | -0.82 | 5.34 E8 |

## 4. TECHNOLOGY ADVANCES

The programmatic gap between the end of phase A and the start of phase B has been used for a number of developments of components, technologies and procedures to enhance the technical readiness of METIS. Some elements that are part of these technology projects are presented elsewhere at this conference, however, a brief selection of the developments is given in this section.

### 4.1 Coronagraphs

METIS will offer coronagraphy in both the L/M and N imaging channels to suppress diffraction effects around bright stars so that faint companions and extended circum-stellar structure can be imaged and characterized. Coronagraphs improve the signal to noise in two ways. Firstly, by reducing the intensity of the stellar halo and therefore reducing the photon shot noise at the planet's location. Secondly, all optical components in the ELT and its instruments have wave front errors that change slowly with time, resulting in so-called Quasi-static Speckles (QSS) in the science camera focal plane, and cause departures from the ideal point spread function (PSF). These QSS are tied to the diffraction halo, and their effect is also reduced with a coronagraph.

**Types of Coronagraphs**
METIS will use at least two types of coronagraphic concepts, the Apodizing Phase Plate (APP), working in the pupil plane, and the Annular Groove Phase Mask (AGPM) in the focal plane. The AGPM [21] provides maximum sensitivity and contrast close in to the star, but, as with all focal plane coronagraphs, it is susceptible to residual tip tilt errors either from the AO system or from the telescope structure. The APP coronagraph [22] (Figure 9, right) does not suffer from this effect, and its robust working design will enable it to work and provide suppression when others cannot.

**Performance Estimates**
To provide an estimate of the performance of METIS with a coronagraph, we used an APP coronagraph as a comparison to direct imaging. The AO corrected phase screens are calculated with the AO simulation package YAO, the coronagraph simulator is written in PDL. The PSFs for the simulation are based on phase screens generated with a time resolution of 1 millisecond in batches of 1000 contiguous AO simulation time steps. These short time steps are necessary to resolve speckle structure in the science camera image that introduces additional noise above that expected from photon shot noise alone. At small angular distances (2 – 6 diffraction widths) this speckle noise is a larger contribution than the photon shot noise. (It typically takes up to 1 second for the speckles to become completely uncorrelated.) After running the PSFs through the coronagraph simulator, they are co-added to form a 1 second image. Many statistically independent short exposures are then combined to form a long exposure image and the variance image. The simulations are monochromatic with the central wavelength of the named filter, with no additional vibration in the ELT system and QSS. Many methods to measure and remove QSS are being developed and implemented in the current generation of high contrast imagers. It is very reasonable to assume that when METIS is on the sky, we will have methods to measure and actively remove QSS on ~1 minute timescales. The results of the APP simulations in comparison to direct imaging are shown in Figure 9.

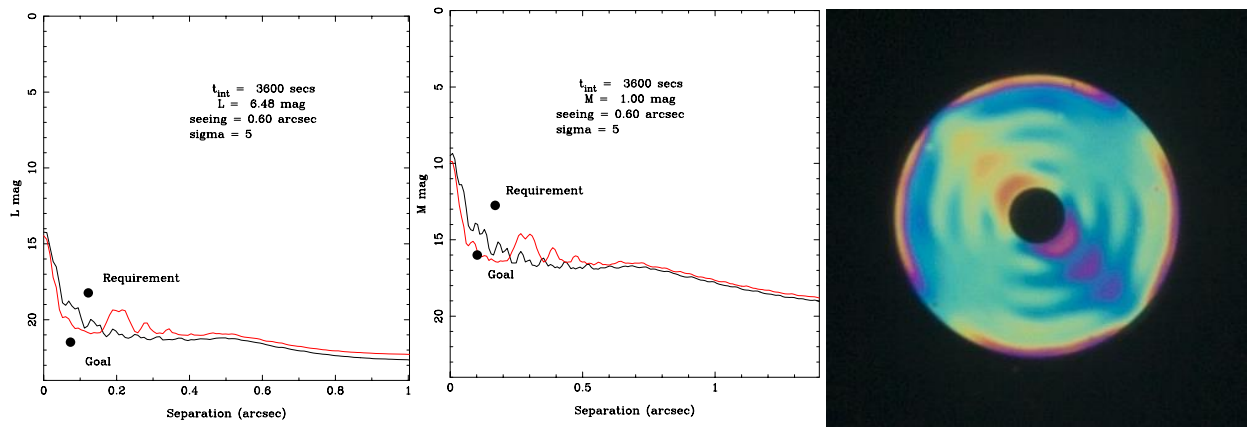

Figure 9 Contrast curves for L (left) and M (center) band imaging with direct imaging (black) and APP Coronagraph imaging (red). Total integration on-sky times of 1 hour and median seeing of 0.60 arcseconds. The requirement and goal values for METIS are also indicated on the plots. Right: prototype Vector APP imaged between two crossed polarizers. It works from 500nm to 800nm and provides a direct imaging contrast of $10^{-4}$.

### 4.2 Immersed Grating

The high resolution spectroscopy that METIS will offer, requires a large format grating. The size of the grating and its surrounding optics can be significantly reduced when the grating is "immersed" in a medium with high index of refraction, like silicon. The benefit of a higher index of refraction concerning the size, turns into a disadvantage with regard to the surface tolerances, which become more critical by the factor of the index of refraction. As previously described [23], we believe that the benefit of the lithography techniques developed by the semi-conductor industry, compensates for the harder requirements, and the initial results look very promising. The grating surface is etched in standard 150 mm wafers, with a groove density of 50 lines/mm. The wafer will be combined by optical bonding with a Si-prism. The very high surface quality allows the surfaces to bond via van der Waals forces. Special tools were developed at SRON, in collaboration with industry, to perform the bonding under vacuum at the required cleanliness and positional accuracy. The shape of the prism, the angles of the immersed grating in the optical system, and specially designed antireflection and absorption coatings will keep straylight and ghosts below the tolerable level. In the fall of 2014, the project will be completed by the optical characterization of the immersed grating demonstrator. Several optical designs are discussed in a separate paper at this conference [24].

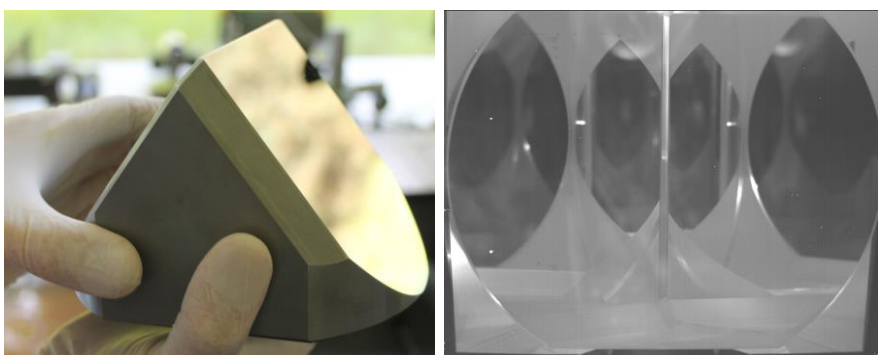

Figure 10 Left: The Si-prism, onto which the prepared wafer will be bonded (bottom surface). Right: infrared photo of the multiple reflections from the surfaces of the material.

### 4.3 Mirror Polishing

The wide operational wavelength range of METIS favors reflective optics. However, the long optical path length and the high spatial resolution of METIS drive the optical performance of the mirrors. The spatial scales necessitate large (60-

300 mm) optical components, and the surface requirements of these mirrors (< 25 nm rms) push the mirror polishing to its limits. In a collaborative effort, MPIA and the Fraunhofer Institut für Angewandt Optik und Feinmechanik (IOF) worked on mirrors made of an Al-alloy, with its coefficient of thermal expansion (CTE) matched to that of NiP, which enables efficient polishing. Given the same CTE, the NiP layer does not degrade the optical surface of the mirror when cooled down to the operating temperature.

The Mirror components have been aged corresponding to the elaborated procedures. For the diamond milling procedure of all mirror substrates a corresponding process development has been applied and tested. For testing of temperature und aging behavior of metal optics, the complete process chain of all polishing techniques has been applied to a plane mirror. After coating with NiP, and iterative diamond milling a correcting MRF technique has been applied for fine tuning the shape. The mirror size is about $56 \times 77$ mm$^2$, and the free aperture of $46 \times 67$ mm$^2$ has a form deviation of only 11 nm rms. Based on the measured reproducible cryogenic error a correcting MRF polishing procedure has been applied, resulting in a final shape error of only 16 nm rms at -190°C. [25]

### 4.4 Coatings

Together with Artemis Optical, UK-ATC studied the technological readiness of the coatings that are required for the entrance window of METIS and the dichroic for the AO pick-off. The following drivers were placed on the AR coating of the *entrance window*: a very high transmission (T > 95%) at science wavelengths (2.9 – 14 µm), also to reduce the thermal emission, and a decent throughput (average T > 70%) at AO WFS wavelengths (0.6 – 2.5 µm). Similar requirements were placed on the *dichroic*: T > 90% from 2.9 – 14 µm, and R > 50% (average) from 0.6 – 2.5 µm. Special consideration was given to wavefront errors and homogeneity across the full aperture.

First results showed that the wavefront errors that were measured were dominated by the substrate flatness; the coating contributed only marginally to the wavefront error. The homogeneity of the coating looked promising over a diameter of 200 mm, suggesting that both window and dichroic can be manufactured to specification and will not limit the performance of METIS.

### 4.5 AO Field Selector

In collaboration with PI (Physik Instruments), MPIA is developing a prototype of a cryogenic field selector. Based on Double-Nexline Actuators (Figure 11), an Al (50mm Ø) mirror could be positioned with an accuracy of 0.75 arcseconds with a setting time ≤ 1 second at cryogenic temperatures (60K < T < 100K). This mirror also enables differential tracking and field derotation for the wavefront sensor pick off.

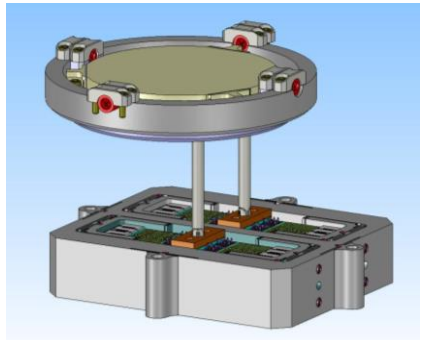

Figure 11 Double- Nexline actuator used together with rod elements for a cardanic mirror mount.

### 4.6 Cryogenic 2D Beam Chopper

Due to the large size and mass of the secondary mirror and the complexity of the primary structure, ELTs will not provide classical chopping and nodding. A METIS Cold Chopper (MCC) demonstrator was realized to show that fast and accurate beam chopping within the METIS cryostat is technical feasible and reliable. The MCC switches the optical beam between the target and a nearby reference sky during observation for characterization of the fluctuating IR background signal in post-processing. On classical telescopes this so called "chopping" is done with the secondary mirror. However, this is not an option for the E-ELT due to the physically large mirror sizes, and alternative solutions had to be developed. The MCC will be located in the pupil of METIS and a cold stop will be integrated, but this is not included in the demonstrator. For the interested reader, the development and characterization of the MCC demonstrator is described in more detail in a separate paper [26].

The MCC mirror (Ø64mm) has to tip/tilt in 2D with a combined angle of up to 13.6 mrad to achieve a chop throw of 8″ on sky. It must be able to reconfigure within 5 ms, to minimize the impact on detector duty cycle (95% at 5 Hz), with a stability and repeatability of 1.7 µrad (3σ), to maintain a highest image contrast. Furthermore, the chop frequency must be variable between 0.1 and 10 Hz. Nonetheless, the chopper must function in the temperature range of 40 to 300 K, consume less than 1 Watt of power at 80 K, and maintain performance during at least 100 million cycles.

The starting point of the MCCD design is its aluminum mirror with an outer diameter of 64 mm and inner diameter of 21 mm. This mirror is constraint in three degrees of freedom (DOF) (x, y, Rz) by a stiff support and the remaining three DOFs, the angular position and focus, are controlled using three linear actuators (custom voice coil actuators) and three interferometers as linear position sensors. In the MCCD, shown in Figure 12, the actuator forces are not directly applied to the mirror. Instead, the forces act on an intermediate body to which the mirror is connected by means of a leaf springs suspension. This stress free support precisely constrains the six DOFs of the mirror and minimizes the impact of intermediate body deformations on the mirror surface accuracy.

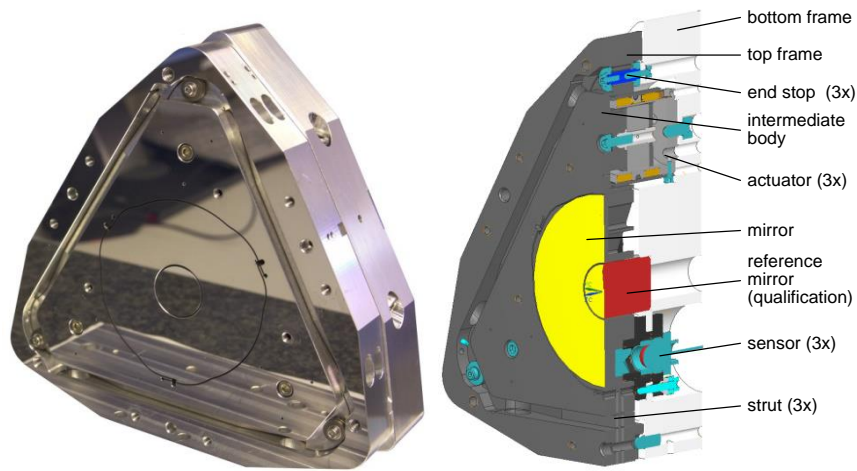

Figure 12 left: photo of the METIS Cold Chopper Demonstrator (MCCD); right: cross-section showing the critical components

The results of our performance tests show that the demonstrator meets most of the MCC requirements. Furthermore, the MCCD has just successfully passed the lifetime test of more than 100 million cycles without noticeable damage or wear. The only remaining issue, the long-term position stability, can likely be solved by reducing the amplifier noise and fine-tuning the controller. The MCC has successfully demonstrated its suitability for METIS.

### 4.7 Drift Scanning

Accurate calibration of mid-infrared observations from the ground is challenging due to the strong and rapidly varying thermal background emission. The classical solution is the chopping/nodding technique. However, chopping is generally inefficient and limited by the mass and size of the secondary mirror in accuracy and frequency. A more elegant solution may be so-called drift scans, where the telescope slowly drifts across or around the region of interest. The source moves on the detector by at least one FWHM of the PSF within the time over which the detector performance or the background emission can be considered stable. The final image will be mathematically reconstructed from a series of adjacent short exposures.

Drift scanning has recently received a lot of interest, mainly for two reasons. First, the new large format mid-IR Si:As detectors (AQUARIUS) suffer from excess low frequency noise (ELFN). In order to reach their nominal performance limit, chopping would have to be performed at a high frequency, faster than what most telescopes can handle. Second, the next generation of extremely large telescopes will not offer chopping/nodding and alternative methods needs to be developed and tested.

**Spatial Gain Variations and Signal-to-Noise**
In the background limited case, the signal to noise ratio (SNR) is determined by the number of photons from the source, the background, and the integration time. However, spatial detector gain variations have an influence on the SNR as they constitute an additional noise term (equation 1). For increasing gain fluctuations, the SNR decreases as shown in Figure 13 (left).

Equation 1

$$SNR = \frac{S\sqrt{t}}{\sqrt{n_p\left(\sigma_b^2 + i^2\sigma_g^2\right)}}$$

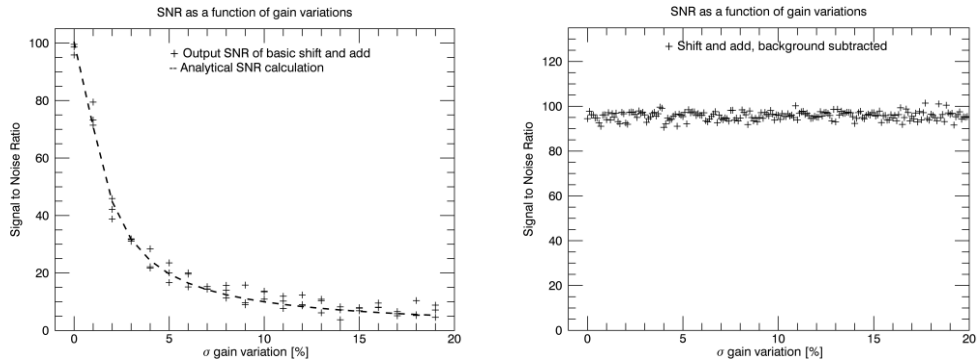

Figure 13 Left: the points are determined with simulated drift scan data with different gain variations. The dashed line is the analytical result; right: SNR as function of gain variation with background subtracted simulated drift scan images.

**Drift Scanning for Background Calibration**
Drift scanning across a point source allows us to determine the background level for each pixel by taking the median of each pixel. In Figure 13 (right) we see that the signal to noise ratio stays almost constant for spatial gain variations. By subtracting the background in this manner, we end up with the gain information still in the source signal. However, adding multiple images and therefore averaging the gain from multiple pixels will average out the gain variations.

If the thermal background is constant in time, we remove the background by drift scanning as described above. The main problem is that this only works for point sources, where each pixel takes more sky than (sky+source) measurements. For extended objects one drift scan sequence may not be large enough to enclose enough sky information. Hence, this type of data reduction with drift scanning will not work for extended objects. Currently we are working on a method that uses the fact that the image function does not change, while the sky emission and the detector properties change [27]. What we observe is $d_x = g_x \times (s + b)$, where $d$ is the observed image at the location $x$ in the array, $g$ is the gain, $s$ is our image function and $b$ is the background signal. Using the drift scan technique we can compare the image function in each image. As the image function is moved across the detector, each pixel sees a different part of the image function. From that one can reconstruct the gain variation and the background signal, even in the presence of time variances. The observer does not need to know what the image function looks like.

## 4.8 Sorption Cooling

METIS is a cryogenically cooled instrument, and the cooling technology is an important issue. Even though the cooling requirements are not driving the design of METIS, vibrations have to be avoided. We have been trying to find out whether astronomical instruments can be cooled without mechanical coolers, and sorption cooling may be a viable solution. Within the context of METIS, the University of Twente has been developing the required technology [28] in collaboration with Dutch Space. The compressors are based on the cyclic adsorption and desorption of a working gas on a sorber material (activated carbon). Cooling is accomplished by adiabatic expansion of the high pressure gas over a flow restriction. The challenge is to raise the cooling power from milli-Watt to Watts. Concerning METIS, a continuous flow liquid nitrogen (LN$_2$) system will provide the ~70K cold stage from where the sorption cooler will cool towards 8K (detector temperature). This is accomplished by three separate cooling chains using three different working gases: Ne, H$_2$, and He.

The study has shown that such a system can be made with present day technologies, complying to the METIS requirements, and fitting within a reasonable volume of the instrument (Figure 14). The next step should be to improve the technology readiness level. Whether this technology can developed and demonstrated in time to be incorporated in METIS is still an open issue.

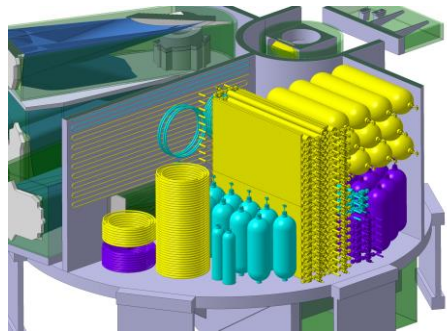

Figure 14 The sorption cooler fitted into METIS. Its space requirement is similar to the high resolution IFU spectrometer. The tubing is not shown all, the "bottles" are pressure tanks, the long pencil-like structures are the compressor cells and the "coils" are the heat exchangers. Yellow represents the He-stage, cyan the H$_2$-stage, and purple the Ne-stage.

## 5. SUMMARY

METIS will be an outstanding first generation instrument on the E-ELT. Focusing on highest angular resolution and high spectral resolution METIS will deliver unique science, complementary to JWST and ALMA, with a strong focus on the science areas of exoplanets, proto-planetary disks and infrared-luminous galaxies. Recent developments on the E-ELT level (size reduction, interface change) did not significantly affect METIS and progress has been made in many areas, including the adaptive optics (AO) concept for METIS. Our technology development program – which ranges from coronagraphic masks, immersed gratings, and cryogenic beam chopper to novel approaches to mirror polishing, background calibration and cryo-cooling – has further enhanced the design and technology readiness of METIS. Both the science case and the technical concept of METIS make it a reliable instrument for early discoveries on the E-ELT in the mid 2020ies.

## REFERENCES

[1] Brandl, B. R. et al., "Instrument concept and science case for the mid-IR E-ELT imager and spectrograph METIS", Proc. SPIE 7735-83 (2010).
[2] Brandl, B. R. et al., "METIS – the thermal infrared instrument for the E-ELT", Proc. SPIE 8446-57 (2012).
[3] Thi et al., A&A 551, 49 (2013)
[4] Blake & Boogert, ApJ 606, 73 (2004)
[5] Salyk et al., ApJ, 731, 130 (2011)
[6] Goto et al., ApJ 652, 758 (2006)
[7] Brittain et al., ApJ 767, 159 (2013)
[8] Pontoppidan et al., ApJ 720, 887 (2010)
[9] http://www.exoplanet.eu/
[10] Casertano et al., A&A 482, 699 (2008)

[11] Lagrange et al., A&A 493L..21L (2009)
[12] Rameau et al., ApJL 772, 15 (2013)
[13] Snellen et al., Nature 465, 1049 (2010)
[14] Brogi et al, Nature 486, 502 (2012)
[15] Bouvier & Wadhwa, Nature Geoscience 3, 637 (2010)
[16] Brandl et al., A&A 543, 61 (2012)
[17] Stuik, R. et al., "Designing the METIS adaptive optics system", Proc. SPIE 8447-131 (2012)
[18] Schmalzl, E. et al., "An end-to-end instrument model for the proposed E-ELT instrument METIS", SPIE 8449-16 (2012).
[19] http://sitedata.tmt.org/index.html
[20] http://www.eso.org/observing/etc/bin/gen/form?INS.MODE=swspectr+INS.NAME=SKYCALC
[21] Mawet, D. et al., "Annular Groove Phase Mask Coronagraph", ApJ 633, 1191 (2005)
[22] Kenworthy, M. et al., "First On-Sky High Contrast Imaging with an Apodized Phase Plate", ApJ 660, 762 (2007)
[23] Van Amerongen, A. H. et al., "Development of silicon immersed grating for METIS on E-ELT", Proc. SPIE 8450-100 (2012)
[24] Agócs et al., "A set of innovative immersed grating based spectrometer designs for METIS", Proc. SPIE, pp. 9151-55 (2014)
[25] Kinast et al., "Minimizing the bimetallic bending for cryogenic metal optics based on electroless nickel", Proc. SPIE, pp. 9151-116 (2014).
[26] Paalvast, S.L. et.al., "Development and characterization of a 2D precision cryogenic chopper for METIS", Proc. SPIE, pp. 9151-12 (2014).
[27] Kuhn, J.R., Lin, H., Loranz, D., "Gain calibrating nonuniform image-array data using only the image data", Publ. Astron. Soc. Pac. 103, 1097 – 1108 (1991)
[28] Ter Brake, H. J. M., et al., "Sorption-based vibration-free cooler for the METIS instrument on E-ELT", Proc. SPIE 8446-295 (2012)